\documentclass{elsart}

\usepackage{natbib}
\usepackage{graphicx}
\usepackage{epsfig}

\usepackage{amssymb}

\begin{document}

\begin{frontmatter}



\title{Detection of CFIRB with AKARI/FIS Deep Observations}


\author[isas]{Woong-Seob Jeong\corauthref{cor}}
\ead{jeongws@ir.isas.jaxa.jp}
\author[isas,esa]{Chris P. Pearson}
\author[snu]{Hyung Mok Lee}
\author[isas]{Shuji Matsuura}
\author[nagoya]{Mitsunobu Kawada}
\author[isas]{Takao Nakagawa}
\author[snu]{Sang Hoon Oh}
\author[isas]{Mai Shirahata}
\author[kasi]{Sungho Lee}
\author[snu]{Ho Seong Hwang}
\author[isas]{Hideo Matsuhara}
\corauth[cor]{Corresponding author. Tel.: +81-42-759-8161; Fax:
+81-42-786-7202.}
\address[isas]{Institute of Space and Astronautical Science, Japan Aerospace
Exploration Agency, Yoshinodai 3-1-1, Sagamihara, Kanagawa 229-8510, Japan}
\address[esa]{ISO Data Centre, European Space Agency, Villafranca del Castillo, P.O. Box 50727, 28080 Madrid, Spain}
\address[snu]{Astronomy Program, Department of Physics and Astronomy, FPRD, Seoul National University,
Shillim-dong, Kwanak-gu, Seoul 151-742, South Korea}
\address[nagoya]{Graduate School of Science, Nagoya University, Furo-cho,
Chigusa-gu, Nagoya 464-8602, Japan}
\address[kasi]{Korea Astronomy and Space
Science Institute, 61-1 Whaam-dong, Yuseong-gu, Daejeon 305-348, South Korea}

\begin{abstract}

The Cosmic Far-Infrared Background (CFIRB) contains information about the
number and distribution of contributing sources and thus gives us an important
key to understand the evolution of galaxies. Using a confusion study to set a
fundamental limit to the observations, we investigate the potential to explore
the CFIRB with \textit{AKARI}/FIS deep observations. The Far-Infrared Surveyor
(FIS) is one of the focal-plane instruments on the \textit{AKARI} (formerly
known as \textit{ASTRO-F}) satellite, which was launched in early 2006. Based
upon source distribution models assuming three different cosmological
evolutionary scenarios (no evolution, weak evolution, and strong evolution), an
extensive model for diffuse emission from infrared cirrus, and instrumental
noise estimates, we present a comprehensive analysis for the determination of
the confusion levels for deep far-infrared observations. We use our derived
sensitivities to suggest the best observational strategy for the
\textit{AKARI}/FIS mission to detect the CFIRB fluctuations. If the source
distribution follows the evolutionary models, observations will be mostly
limited by source confusion. We find that we will be able to detect the CFIRB
fluctuations and that these will in turn provide information to discriminate
between the evolutionary scenarios of galaxies in most low-to-medium cirrus
regions.

\end{abstract}

\begin{keyword}
Space-based infrared telescopes \sep Far infrared \sep Infrared cirrus \sep
Evolution of extragalactic objects

\PACS 95.55.Fw \sep 95.85.Gn \sep 98.58.Ca \sep 98.62.Ai
\end{keyword}
\end{frontmatter}


\section{Introduction}

Measurements of the infrared background have been carried out in several
wavelength bands. The study of deep galaxy counts at infrared wavelengths has
enabled us to resolve a significant fraction of the Cosmic Far-Infrared
Background (CFIRB) which contains the accumulated intensity from redshifted and
unresolved extragalactic sources. In addition to point sources, there also
exist fluctuations in the surface brightness of extended structure due to
neutral interstellar dust in the Milky Way that is heated by the interstellar
radiation field, known as the infrared cirrus \citep{low84}. Moreover, although
significant fluctuations from zodiacal emission have not yet been detected
\citep{abra97}, this component may also contribute to the background
fluctuations. Cirrus emission peaks at far-IR wavelengths but was detected in
all 4 \textit{IRAS} bands. The infrared emission from Galactic cirrus is
usually a function of Galactic latitude and is serious for wavelengths longer
than 60 $\mu$m. In many cases, the spatial power spectrum from fluctuations of
the dust emission can be expressed as a simple power-law with the power index
of $\sim$-3 \citep{gautier92}. Unresolved sources below the detection limit
also create fluctuations in the background, which was indeed detected from
Infrared Space Observatory (ISO) observations \citep{lagache00,matsuhara00}. By
using the different properties of spatial fluctuations compared to unresolved
sources and cirrus emission, some authors \citep{kiss01,miville02,gross06} have
obtained significant levels of CFIRB fluctuation. However, since a dependency
of the power index of the cirrus power spectrum on both the wavelength and
surface brightness is suggested  from recent observations
\citep{kiss03,ingalls04}, we need to understand the spatial property of cirrus
fluctuations as a function of Galactic latitude.

The Japanese \textit{AKARI} (formerly known as \textit{ASTRO-F}) satellite was
launched on February 21st 2006 and is performing an all-sky survey
\citep{naka01,shib04,cpp04,cpp07a} and deep pointed observations at the North
and South Ecliptic Poles \citep{matsuhara06} and other selected areas in 4
far-IR bands to much improved sensitivities, spatial resolutions and wider
wavelength coverage than the previous \textit{IRAS} survey over two decades
ago. These far-IR observations will be expected to detect the CFIRB via
fluctuation analysis. The actual observing plan may change depending on the
in-flight performance of \textit{AKARI}. The purpose of this paper is to show
the potential possible detection of the CFIRB and its expected fluctuation
level.

\section{Source Counts Model and Confusion}

Source counts results in the far-IR regime from \textit{ISO} and
\textit{Spitzer} observations \citep{matsuhara00,puget99,dole04} show a
significant evolution in the extragalactic population. The source count models
used in this paper consider two evolutionary scenarios for galaxies revised
from our previous luminosity evolution model \citep{cpp96} and burst evolution
model \citep{cpp01}. For specific updated evolutionary parameters, see
\citet{cpp05,cpp07c}. In addition, a brief explanation and the color analysis
for the revised models are described in \citet{cpp07b}.

\begin{figure*}[ht]
  \begin{center}
    \epsfxsize = 6.8cm
    \epsffile{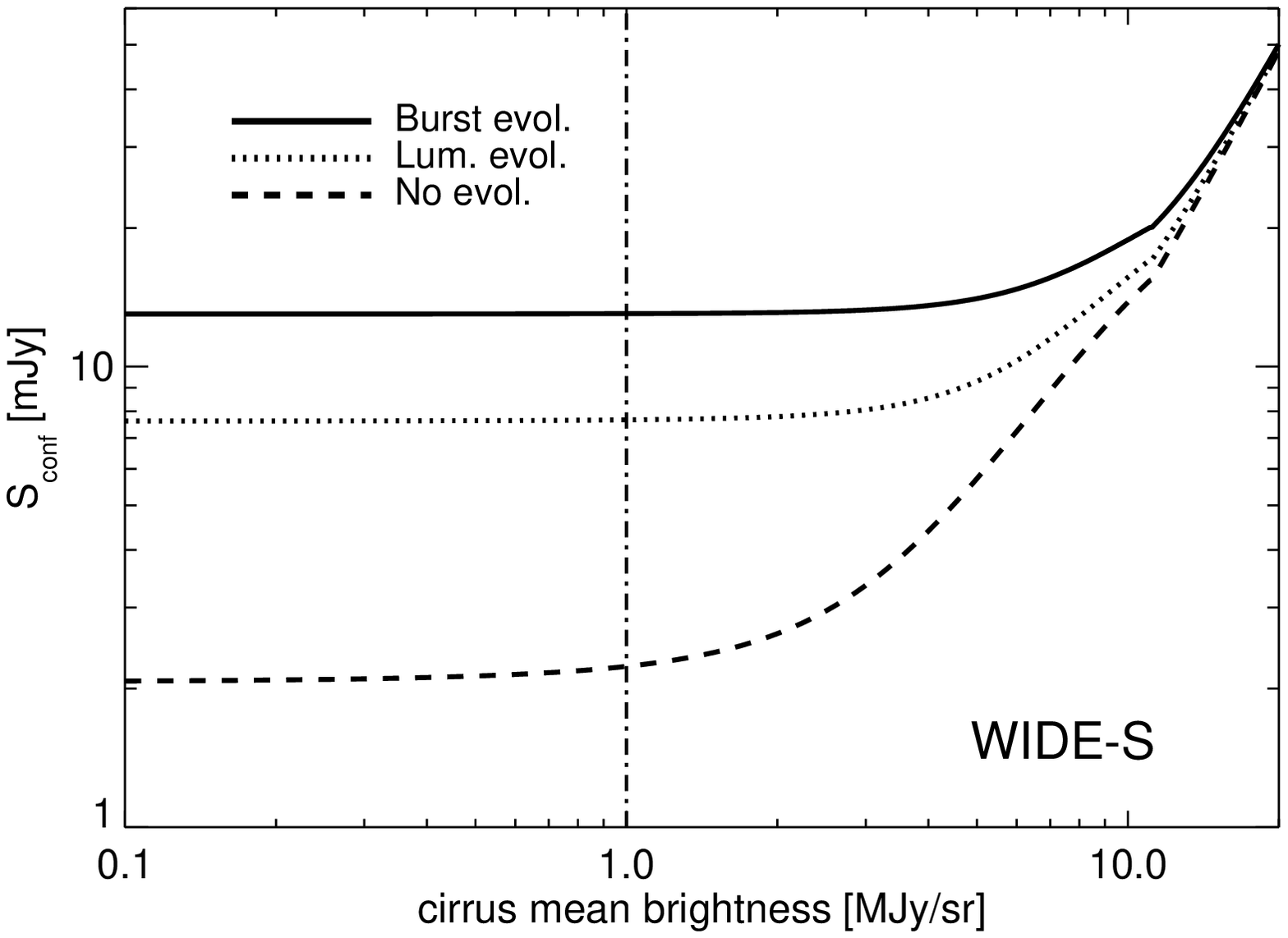}
    \epsfxsize = 6.8cm
    \epsffile{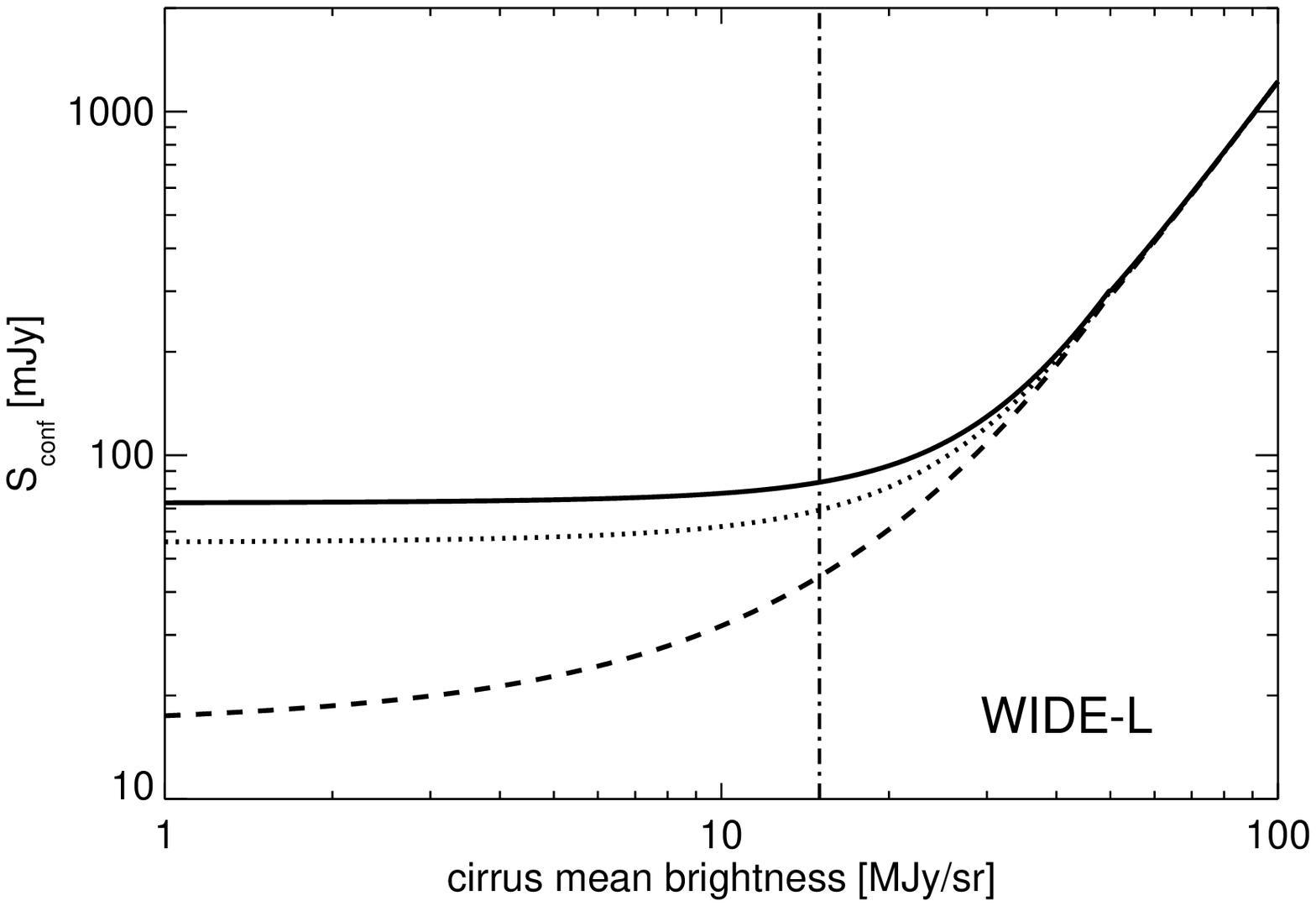}
    \end{center}
   \caption{Confusion limits considering both sky confusion and source confusion
   from careful simulations for the \textit{AKARI} mission.
   The two vertical lines show the mean cirrus brightness 1.0 MJy~sr$^{-1}$
   for the WIDE-S band (left) and 15 MJy~sr$^{-1}$ for the WIDE-L band (right),
   respectively. The leftwards and rightwards in the figure mean low and
   high cirrus regions, respectively. In low-level cirrus regions, source confusion
   is expected to dominate. Note that sky confusion increases as the mean cirrus
   brightness becomes larger while the source confusion has a constant value
   irrespective of the mean cirrus brightness.}
   \label{fig_final_conf}
\end{figure*}

Due to the fluctuations from background structure and unresolved sources  below
the resolution of the telescope beam, the detection of sources is usually
influenced by confusion effects. The sky confusion due to cirrus structures
depends on sky position, however the source confusion due to unresolved sources
remains constant everywhere except for clustering effects of sources and solar
system objects. Since the cirrus fluctuations are represented by near-Gaussian
noise, the final confusion limits $S_{\rm conf}$ can be approximated by the
summation of the two confusion noise components:
\begin{equation}
S_{\rm conf}(\mathbf{r}) = \sqrt{S_{\rm cc}(\mathbf{r})^2 + S_{\rm sc}^2},
\label{eq_tot_no}
\end{equation}
where $S_{\rm cc}$ is the sky confusion limit due to cirrus structure, $S_{\rm
sc}$ is the source confusion limit and $\mathbf{r}$ is the sky position. The
source confusion estimated with the two evolutionary models of
\citet{cpp05,cpp07c} is 7.6 $-$ 13 mJy and 55 $-$ 72 mJy respectively (lower
and upper limit is for luminosity evolution and burst evolution model,
respectively), for the two wide bands of \textit{AKARI} mission, i.e., WIDE-S
and WIDE-L band with central wavelengths of 90 $\mu$m and 140 $\mu$m. The sky
confusion can be scaled with the cirrus surface brightness
\citep{HB90,kiss01,kiss05,jeong05} and fig. \ref{fig_final_conf} shows the
final confusion limits for an assumed range of cirrus brightness levels for the
\textit{AKARI} mission.

\section{Expected Cosmic Far-Infrared Background}

The flux levels from a randomly distributed population of extragalactic sources
below the detection limit create fluctuations in the observed background. The
CFIRB intensity $I_{\rm CFIRB}$ and fluctuation $P_{\rm CFIRB}$ produced by all
sources with a flux below the limiting flux $S_{\rm lim}$, are obtained from:
\begin{equation}
  I_{\rm CFIRB} = \int^{S_{\rm lim}}_{0} S~{dN \over dS}~dS
\label{eqn_int_cfirb}
\end{equation}

and

\begin{equation}
  P_{\rm CFIRB} = \int^{S_{\mathrm{lim}}}_{0} S^2~{dN \over dS}~dS,
\label{eqn_fluc_cfirb}
\end{equation}
where $S$ is the flux and $dN/dS$ are the differential source counts. The
detection limits $S_{\mathrm{lim}}$ for \textit{AKARI} estimated from our study
including both sky confusion and source confusion \citep{jeong05,jeong06} with
our revised source counts models \citep{cpp05,cpp07c} are shown in Figure
\ref{fig_final_conf}. In the dark cirrus fields, the confusion limits are
expected to be 7.7 -- 13 mJy for the WIDE-S band and 56 -- 73 mJy for the
WIDE-L band (lower and upper values are for luminosity and burst evolution
models, respectively).

\begin{figure}[ht]
  \begin{center}
    \epsfxsize = 6.8cm
    \epsffile{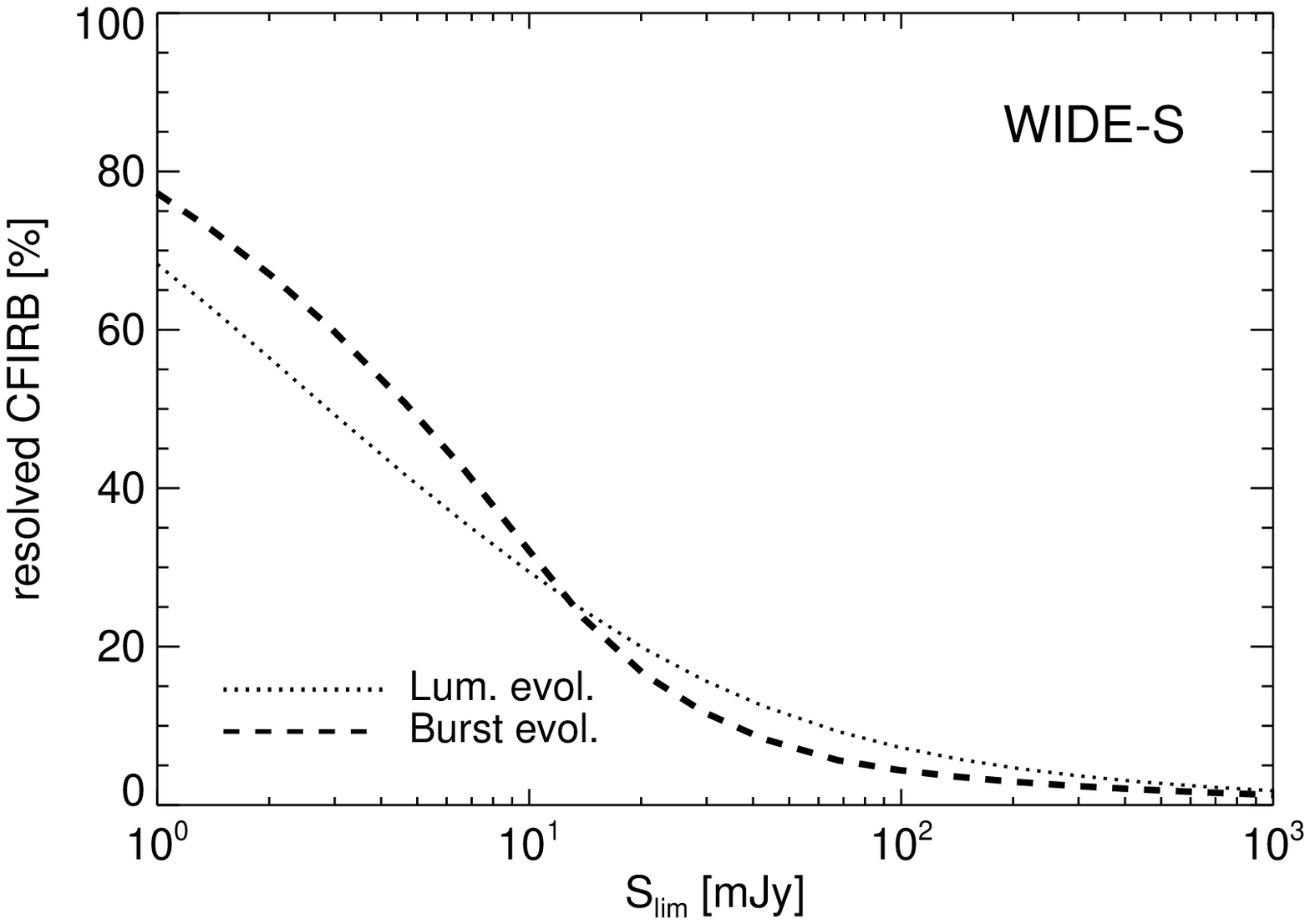}
    \epsfxsize = 6.8cm
    \epsffile{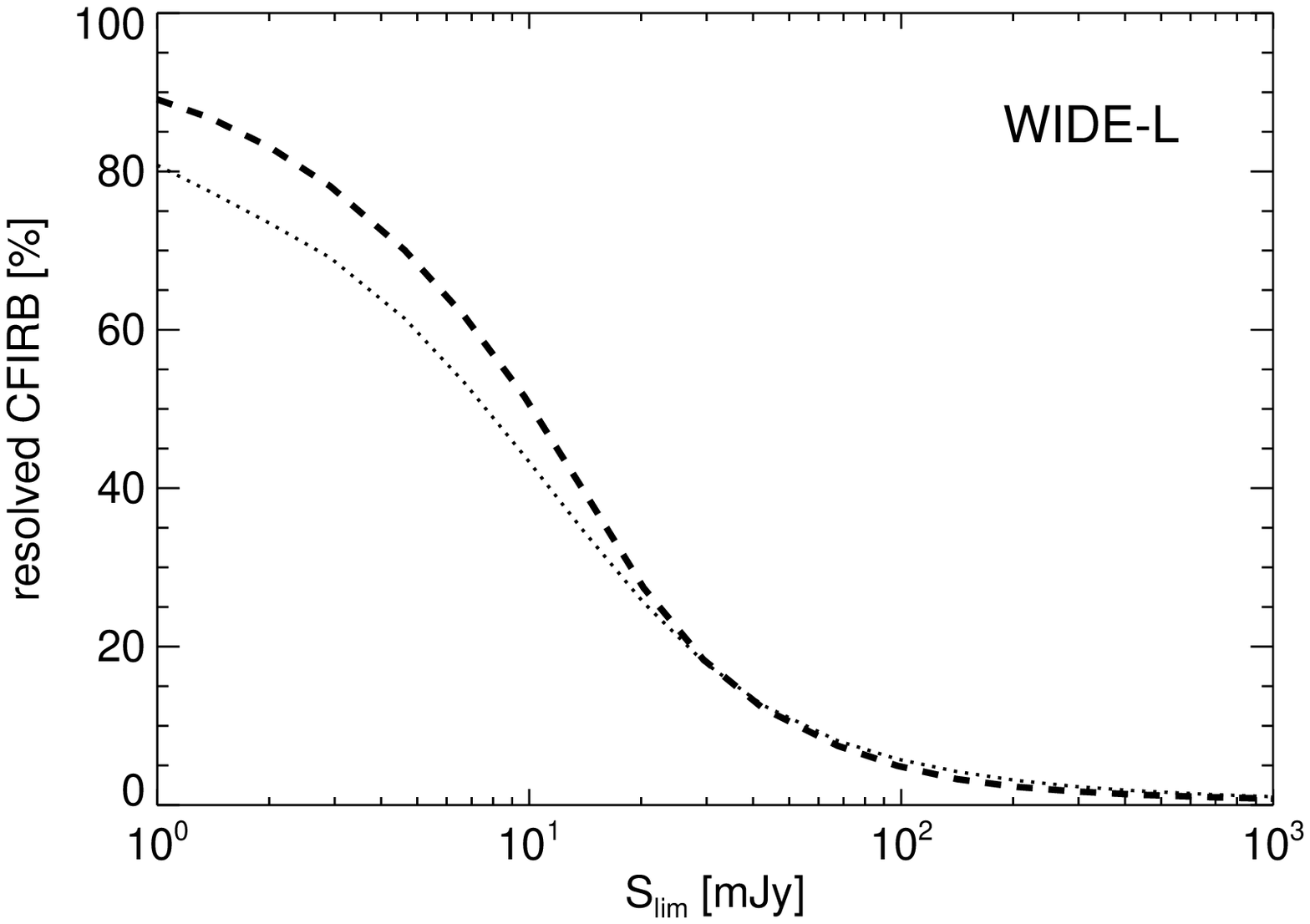}
    \end{center}
   \caption{Fraction of resolved CFIRB for the WIDE-S and WIDE-L bands.
   $S_{\mathrm{lim}}$ is the limiting sensitivity.}
   \label{fig_resol_cfirb}
\end{figure}

\begin{figure}[ht]
  \begin{center}
    \epsfxsize = 6.8cm
    \epsffile{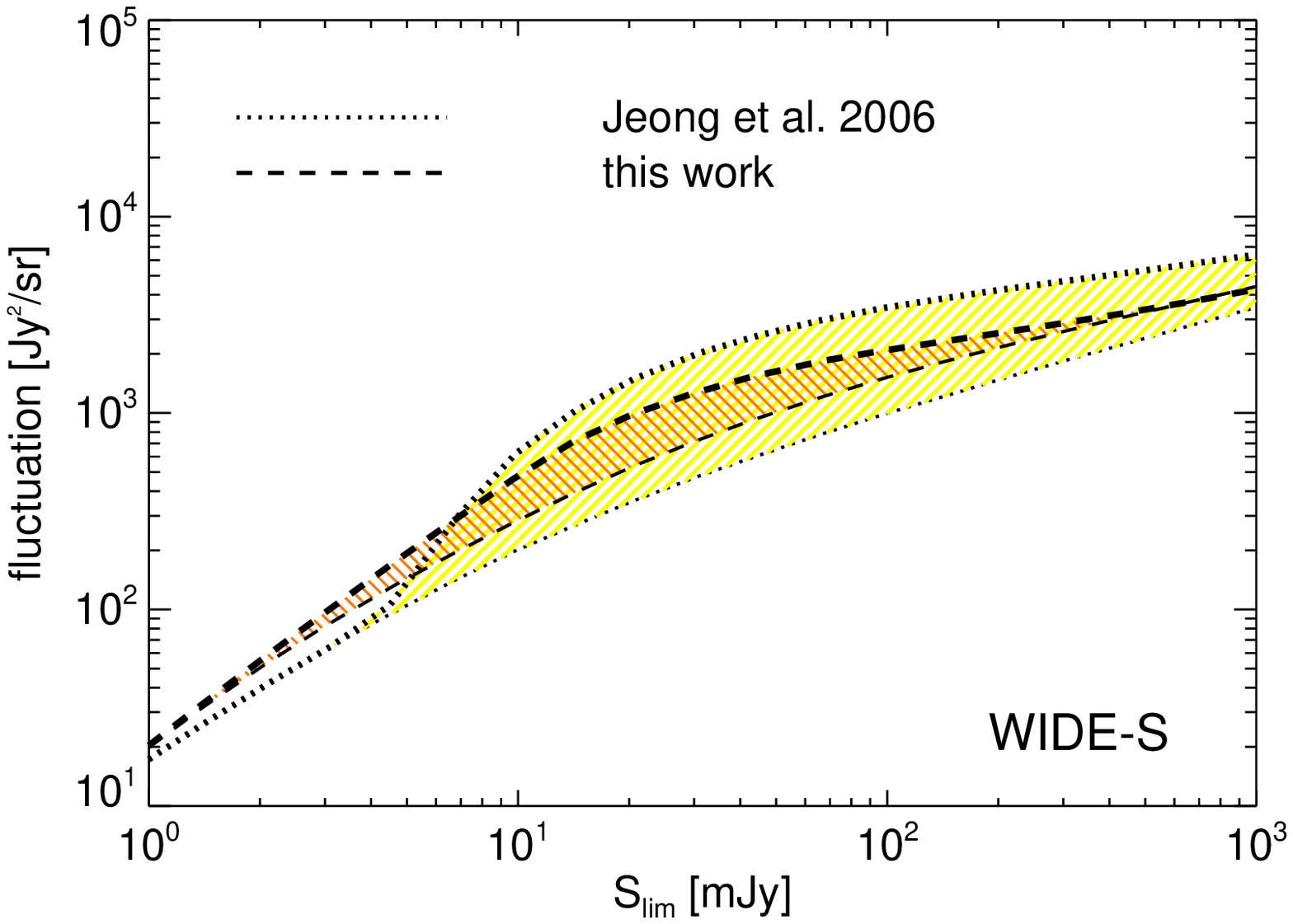}
    \epsfxsize = 6.8cm
    \epsffile{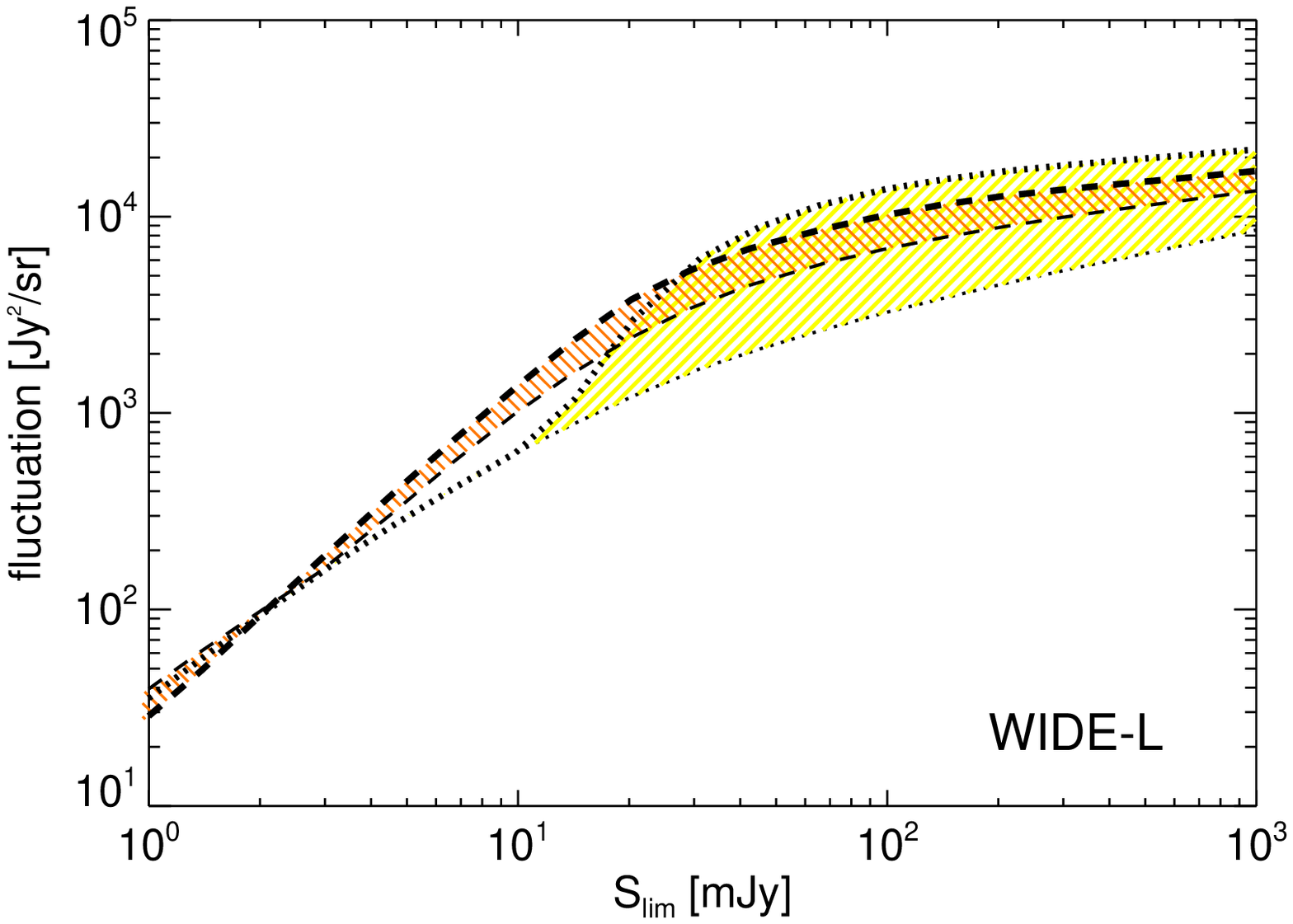}
    \end{center}
   \caption{Expected CFIRB and cirrus fluctuations in the WIDE-S and WIDE-L bands
   for the \textit{AKARI} mission. The shaded area covers fluctuations from
   both evolutionary models. For comparison, we show the results for our revised and
   previous models.
   The lower limit is for the luminosity evolution model and the upper limit for burst
   evolution model.}
   \label{fig_fluc_cfirb}
\end{figure}

Due to our better sensitivity, we can resolve more sources from the CFIRB
compared to previous missions. Figs. \ref{fig_resol_cfirb} and
\ref{fig_fluc_cfirb} show the amount of the resolved CFIRB and the CFIRB
fluctuation for the WIDE-S and WIDE-L bands. The amount of resolved CFIRB
increases  faster (i.e., decreasing of CFIRB fluctuation) in the WIDE-L band
compared with that in the WIDE-S band. We list the expected CFIRB intensity,
fluctuation and resolved fraction of CFIRB in Table \ref{tab_int_cfirb}. The
expected 5$\sigma$ sensitivity for a diffuse source in \textit{AKARI}'s deep
imaging mode (slow scan in 15 arcsec/sec, 2 sec reset) is 0.34 and 0.56 MJy/sr
for the WIDE-S and WIDE-L bands respectively \citep{verdugo07}. Though the
CFIRB intensity expected in both evolutionary models is still below 3$\sigma$
sensitivity of the WIDE-S band, the CFIRB fluctuation in the burst evolution
model is expected to be stronger than that of the noise ($\sim$ 600 Jy$^2$/sr
at 5$\sigma$). However, the detector characteristics may also affect the
accurate estimation of the fluctuation \citep{jeong04,shirahata04}. In the
far-IR range, \citet{lagache00} and \citet{matsuhara00} have studied the
detection of the CFIRB fluctuation from ISO observations in the Marano 1 field
and the Lockman Hole, respectively. For comparison, we list our estimated
fluctuations in Table \ref{tab_fluc_comp}. We find that the observed
fluctuations are mostly located between the results predicted from our two
evolutionary models except for the observations at 90 $\mu$m. Comparing with
the model of \citet{lagache03}, our estimated fluctuations from the burst
evolution model are in good agreement. Since the deep observations of
\textit{AKARI} will have better sensitivity and resolution over a wider sky
area than the \textit{ISO} observations, we expect that we will be able to
obtain more accurate fluctuation levels from our observations.

\begin{figure}[ht]
  \begin{center}
    \epsfxsize = 6.8cm
    \epsffile{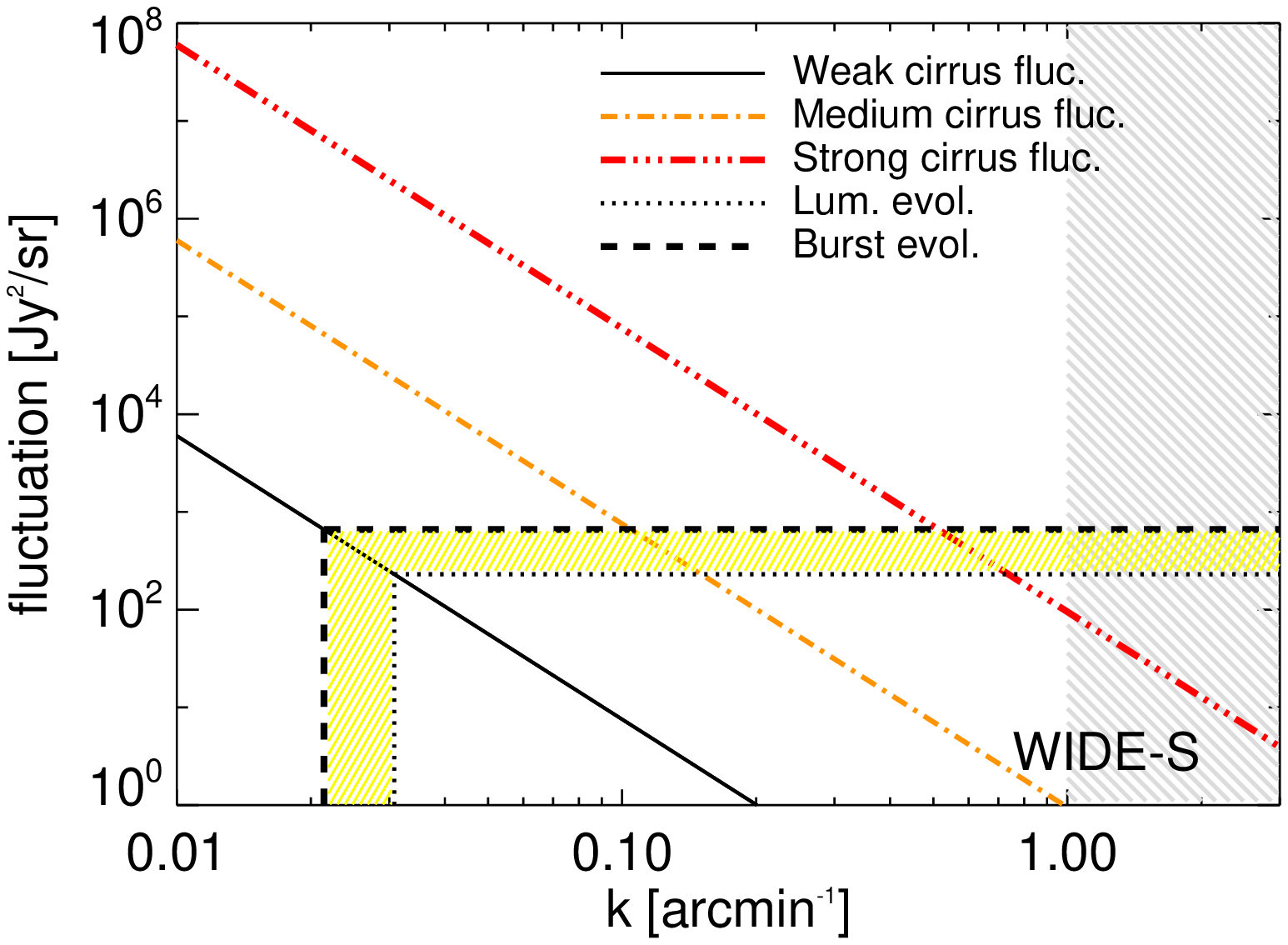}
    \epsfxsize = 6.8cm
    \epsffile{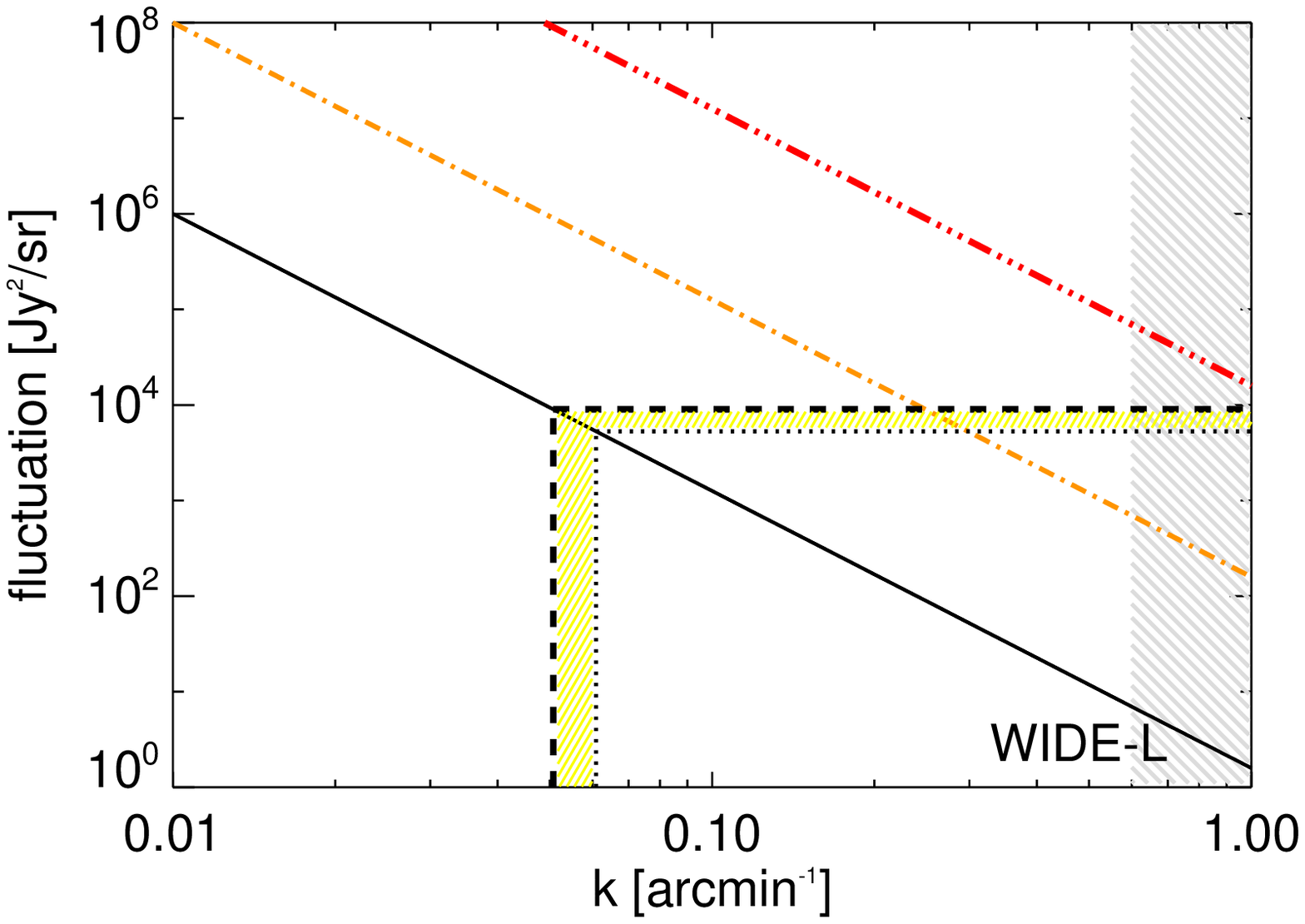}
    \end{center}
   \caption{Expected CFIRB and cirrus fluctuations in the WIDE-S and WIDE-L bands
   for \textit{AKARI}.
   The solid line shows the power spectrum of the cirrus emission at high Galactic
   latitude and the shaded area covers the fluctuations from the two evolution models.
   The lower limit is for the luminosity evolution model and the upper limit for
   the burst evolution model. The shaded area to the right means that those
   regions should be limited by the spatial resolution of the \textit{AKARI} mission.}
   \label{fig_ps_cfirb}
\end{figure}

The estimation of fluctuations from the analysis of spatial power spectrum is
usually used in the analysis of the CFIRB due to the lack of necessity for
accurate absolute calibration. The power spectrum of cirrus emission at high
Galactic latitudes ($>$ 80 degree) also has a fluctuation of around 10$^6$
Jy$^2$/sr at 0.01 arcmin$^{-1}$ and 160 $\mu$m and a power index of -2.9 $\pm$
0.5. The shaded area in Figure \ref{fig_ps_cfirb} covers the fluctuations
predicted by both evolutionary models. In the WIDE-L band, we may require a
minimum areal coverage of $\sim$ 280 $-$ 400 arcmin$^2$ in order to effectively
distinguish the CFIRB fluctuations in the observed data that will inevitably
include both CFIRB and cirrus emission. If we use the WIDE-S band, the minimum
required area coverage increases by a factor of $\sim$ 5.

\begin{figure}[ht]
  \begin{center}
    \epsfxsize = 6.8cm
    \epsffile{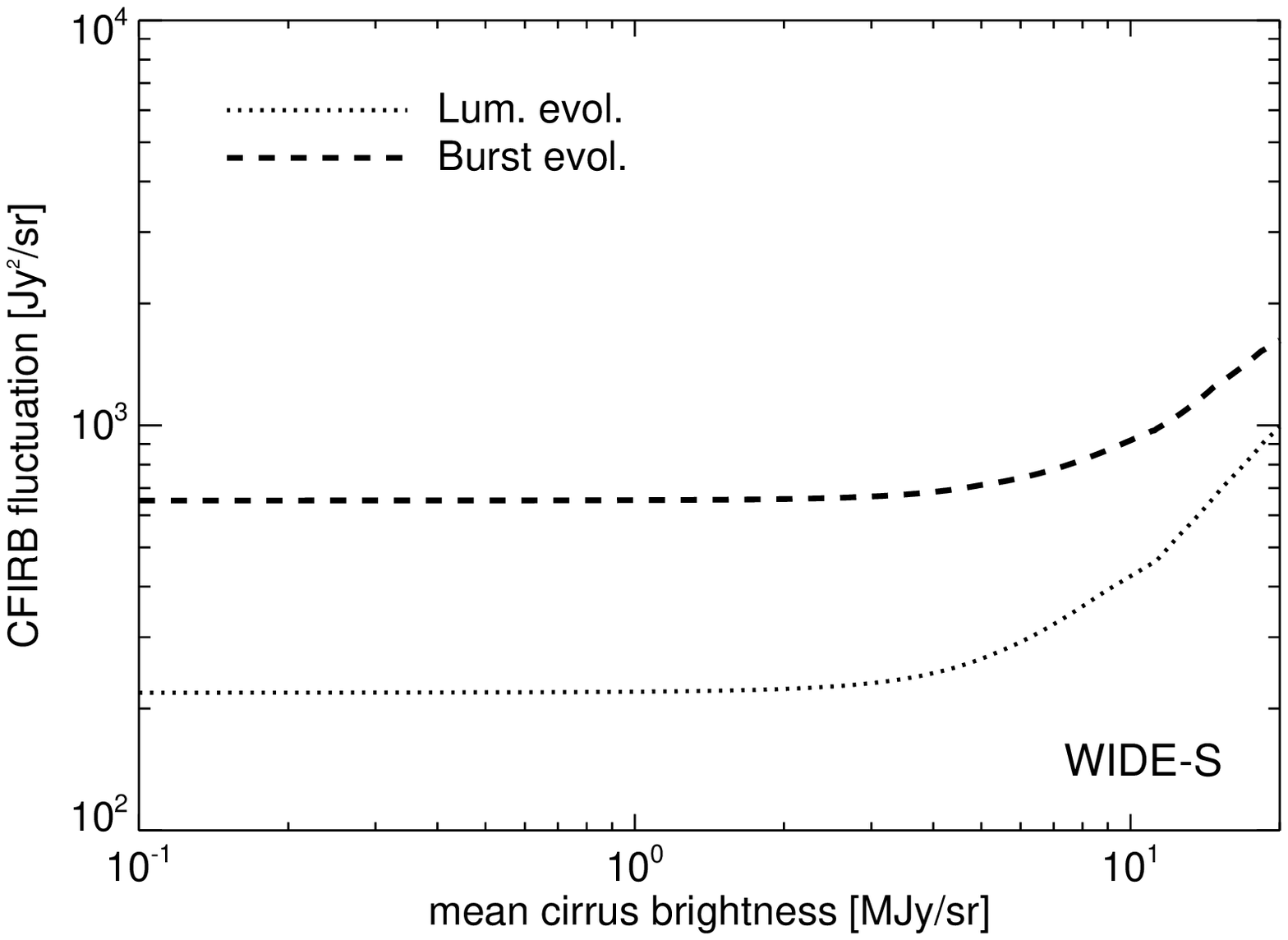}
    \epsfxsize = 6.8cm
    \epsffile{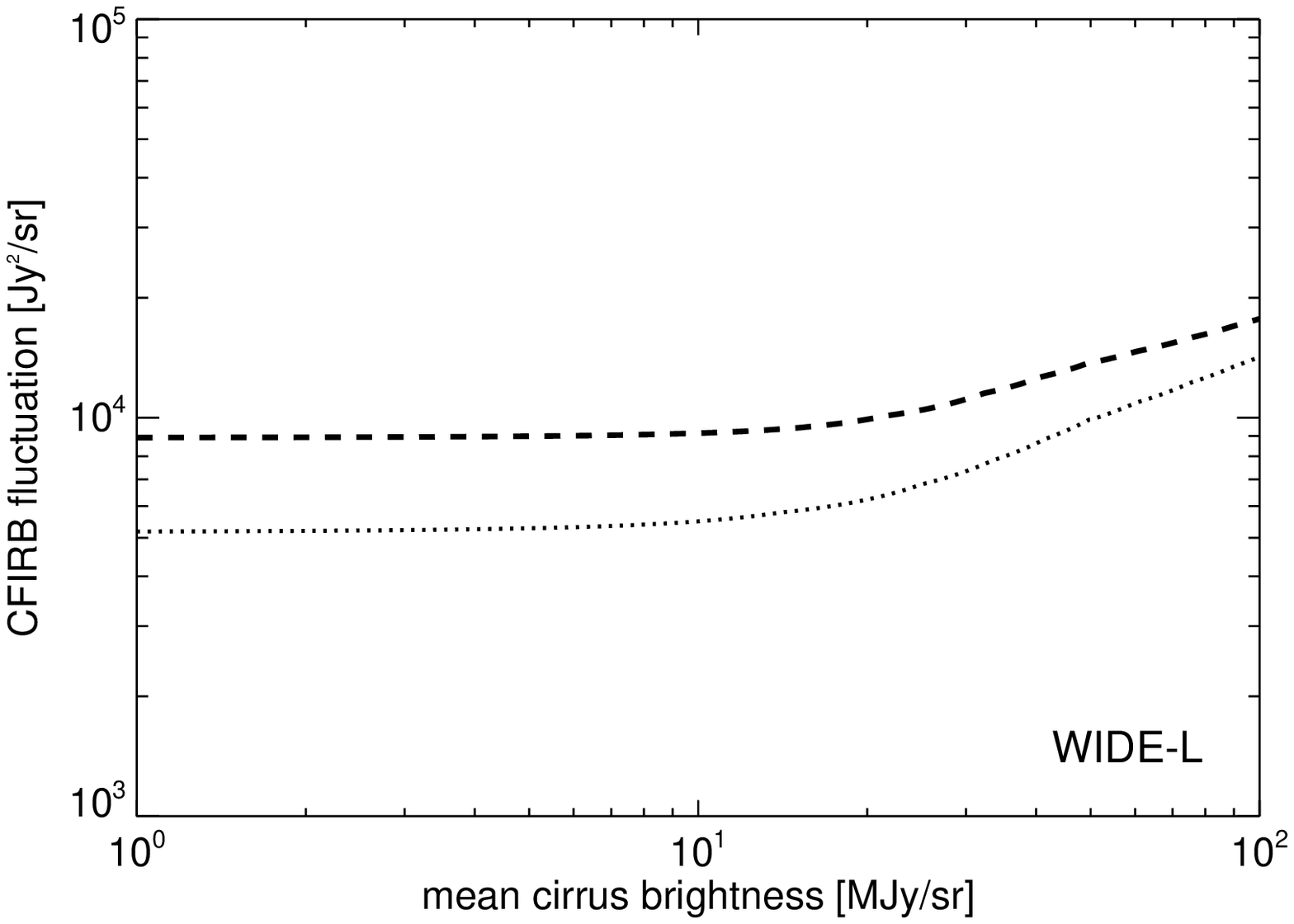}
    \end{center}
   \caption{Expected CFIRB fluctuation as a function of mean cirrus brightness
   in the WIDE-S and WIDE-L bands for the \textit{AKARI} mission. The CFIRB
   fluctuation shows a monotonic increase in the medium mean cirrus brightness
   regions as the mean cirrus brightness increases.}
   \label{fig_cfirb_detlim}
\end{figure}

Obviously, as we observe higher cirrus regions, the detection limit becomes
worse due to the increasing sky confusion noise, while the CFIRB fluctuation
becomes larger (see the Figure \ref{fig_fluc_cfirb}). As can be seen in Figure
\ref{fig_cfirb_detlim}, the fluctuations show a monotonic increase to medium
cirrus regions. In the case of low Galactic latitude regions, since the cirrus
fluctuation power should be more than 10$^{8}$ (WIDE-S) and 10$^{10}$ (WIDE-L)
Jy$^2$/sr at 0.01 arcmin$^{-1}$ (see Figure \ref{fig_ps_cfirb}), it will be
difficult to extract the CFIRB fluctuation from the analysis of the power
spectrum (Note that this CFIRB fluctuation has a spatial scale close to the
resolution of the detector). However, we expect to detect CFIRB fluctuations in
most low-to-medium cirrus regions.

\section{Summary}

While the CFIRB fluctuation is expected to be almost constant irrespective of
position on the sky, except for low Galactic latitude regions (due to severe
sky confusion), cirrus fluctuations should be different depending on the
regions of sky. We verified these results by comparing our simulated results
with observational data from ISO \citep{kiss01,kiss05,jeong05}. Thus, in order
to estimate CFIRB fluctuations effectively, we need to compare the cirrus
fluctuation via observations of regions at different Galactic latitudes (cirrus
brightness). The deep observations planned with the \textit{AKARI} mission will
enable us to detect the CFIRB fluctuation in most regions of low-to-medium
cirrus brightness in both the WIDE-S and WIDE-L bands. These observations will
lead to a better understanding of the Cosmic Far-Infrared Background (CFIRB) -
an important key to reveal the evolutionary scenario of galaxies in Universe.


\section*{ACKNOWLEDGEMENTS}

Woong-Seob Jeong acknowledges a Japan Society for the Promotion of Science
(JSPS) fellowship to Japan.

\newpage

\begin{table}[h]
\centering \caption{Expected intensity, fluctuation and resolved fraction of
CFIRB.} \label{tab_int_cfirb}
\begin{center}
\begin{tabular}{ccccc} \hline
& \multicolumn{2}{c}{Luminosity evolution} & \multicolumn{2}{c}{Burst
evolution} \\
 & WIDE-S & WIDE-L & WIDE-S & WIDE-L
 \\\hline
 Intensity (MJy/sr) & 0.12 & 0.50 & 0.17 & 0.66 \\
 Fluctuation (Jy$^2$/sr) & 230 & 5300 & 660 & 9000  \\
 Potential resolution (\%) & 33 & 10 & 26 & 7
\\ \hline
\end{tabular}
\end{center}
\end{table}

\vskip 12pt

\begin {table}[h]
\centering \caption{Comparison of CFIRB fluctuations.} \label{tab_fluc_comp}
\begin{center}
\begin{tabular}{@{}ccccccc}
\hline
$\lambda$ & $\theta$ & $S_{\rm max}$ & Observations & Predicted $^a$ & Predicted $^b$ & Predicted (this work) $^c$\vspace{5pt} \\
 ($\mu$m) & (arcmin) & (mJy) & (Jy$^2$/sr) & (Jy$^2$/sr) & (Jy$^2$/sr) & (Jy$^2$/sr)
\\\hline
 90 & 0.4 $-$ 20 & 150 & 13000 $\pm$ 3000~$^d$ & 5300 & 2100 $-$ 7200 & 3200
$-$ 4700
\\ 170 & 0.6 $-$ 4 & 100 & 7400~$^e$ & 12000 & 3800 $-$ 14000 & 8200 $-$ 13000
\\ 170 & 0.6 $-$ 20 & 250 & 12000 $\pm$ 2000~$^d$ & 16000 & 5500 $-$ 18000 & 11000 $-$ 16000
\\ \hline
\end{tabular}
\medskip
\begin{flushleft}
{\small {\em $^a$} Model from \citet{lagache03}. \\
{\em $^b$} Lower and upper limits from our luminosity and burst evolution
model, respectively \citep{jeong06}. \\
{\em $^c$} Lower and upper limits from modified evolution
models. \\
{\em $^d$} \citet{matsuhara00} \\
{\em $^e$} \citet{lagache00} }
\end{flushleft}
\end{center}
\end{table}

\vskip 12pt


\end{document}